\documentclass[12pt]{article}
\usepackage{diagbox}
\usepackage{mathtools}
\usepackage{bbm}
\usepackage{amsmath}
\allowdisplaybreaks

\mathtoolsset{showonlyrefs}

\renewcommand{\emph}[1]{\textit{#1}}
\usepackage{enumerate,amsmath,amsthm,latexsym,amssymb}
\usepackage{color}\usepackage{graphicx}

\def\bse{\boldsymbol{\epsilon}}

\definecolor{brown}{cmyk}{0, 0.72, 1, 0.45}
\definecolor{grey}{gray}{0.5}

\newcommand{\old}[1]{}

\newcounter{rot}

\newcommand{\ignore}[1]{}

\def\cA{{\mathcal A}}
\def\cB{{\mathcal B}}

\newcommand{\set}[1]{\left\{#1\right\}}

\def\cQ{\mathcal{Q}}

\newcommand{\proofend}{\hspace*{\fill}\mbox{$\Box$}\\ \medskip\\ \medskip}

\def\ii_(#1,#2){i_{#1}^{#2}}

\def\a{\alpha}

\def\d{\delta}
\def\D{\Delta}
\def\e{\varepsilon}

\def\G{\Gamma}

\def\th{\theta}

\def\m{\mu}

\def\1{{\bf 1}}
\def\0{{\bf 0}}

\def\cN{\mathcal{N}}

\newcommand{\brac}[1]{\left( #1 \right)}

\def\E{{\bf E}}

\renewcommand{\Pr}{\operatorname{\bf Pr}}
\newcommand\bfrac[2]{\left(\frac{#1}{#2}\right)}

\newtheorem{theorem}{Theorem}[section]

\newtheorem{lemma}[theorem]{Lemma}
\newtheorem{corollary}[theorem]{Corollary}

\newtheorem{remthm}[theorem]{Remark}

\newenvironment{remark}{\begin{remthm}\rm }{\end{remthm}}%
\newcounter{thmtemp}
\newtheorem{definition}{Definition}

\newcommand{\nospace}[1]{}

\def\path{\operatorname{PATH}}

\newcommand{\beq}[2]{\begin{equation}\label{#1}#2\end{equation}}

\begin{document}
\title{{\bf Randomly coloring simple hypergraphs with fewer colors}}
\author{Michael Anastos
\ \ \ \ Alan Frieze\thanks{
Supported in part by NSF grant DMS-1362785.} \\
Department of Mathematical Sciences,\\ Carnegie Mellon University,\\
Pittsburgh PA15213.\\
{\footnotesize Email: \ manastos@andrew.cmu.edu; alan@random.math.cmu.edu;}
}
\maketitle
\begin{abstract}
We study the problem of constructing a (near) uniform
random proper $q$-coloring of a simple $k$-uniform hypergraph
with $n$ vertices and maximum degree $\D$.
(Proper in that no edge is mono-colored and simple in that two edges
have maximum intersection of size one).
We show that if $q\geq \max\set{C_k\log n,500k^3\D^{1/(k-1)}}$ then the Glauber Dynamics will become close to uniform in $O(n\log n)$ time, given a random (improper) start. This improves on the results in Frieze and Melsted \cite{FM}.
\end{abstract}

\section{Introduction}
Markov Chain Monte Carlo (MCMC) is
an important tool in sampling from complex distributions. It has been
successfully applied in several areas of Computer Science, most
notably for estimating the volume of a convex body \cite{DFK},
\cite{KLS}, \cite{LV}, \cite{CV} and estimating the permanent of a non-negative matrix \cite{JSV}.

Generating a (nearly) random $q$-coloring of
a $n$-vertex graph $G=(V,E)$ with maximum degree
$\Delta$ is a well-studied problem in Combinatorics \cite{BW}
and Statistical Physics \cite{SS}.  Jerrum \cite{Jerrum} proved that a simple, popular
Markov chain, known as the {\em Glauber dynamics}, converges to a random
$q$-coloring after $O(n\log{n})$ steps, provided $q/\D>2$.
This led to the challenging problem of determining the smallest
value of $q/\Delta$ for which a random $q$-coloring can be generated in
time polynomial in $n$. Vigoda \cite{vigoda} gave the first significant
improvement over Jerrum's result,
reducing the lower bound on $q/\D$ to $11/6$ by analyzing a different Markov chain.
There has been no success in extending Vigoda's approach to smaller values of $q/\D$,
and it remains the best bound for general graphs. There are by now several papers giving improvements
on \cite{vigoda}, but in special cases. See Frieze and Vigoda \cite{FriVig} for a survey.

In this paper we consider the related problem of finding a random coloring
of a simple $k$-uniform hypergraph.
A $k$-uniform hypergraph $H=(V,E)$ has vertex set $V$ and $E=\set{e_1,e_2,\ldots,e_m}$
are the edges. Each edge is a $k$-subset
of $V$. Hypergraph $H$ is simple if $|e_i\cap e_j|\leq 1$ for $i\neq j$. A coloring
of $H$ is proper if every edge contains
two vertices of a different color. The chromatic number $\chi(H)$ is the smallest
number of colors in a proper coloring of $H$.
In the case of graphs $k=2$ we have $\chi(G)\leq \D+1$ but for hypergraphs ($k\geq 3$)
we have much smaller bounds. For example
a simple application of the local lemma implies that $\chi(H)=O(\D^{1/(k-1)})$. In fact
a result of Frieze and Mubayi \cite{FM1},
is that for simple hypergraphs $\chi(H)=O((\D/\log\D)^{1/(k-1)})$. The proof
of \cite{FM1} is somewhat more involved.
It relies on a proof technique called the ``nibble``. 
The aim of this note is to show how to improve the results of Frieze and Melsted \cite{FM} who showed that under certain circumstances simple hypergraphs can be efficiently randomly colored when there are fewer than $\D$ colors available. In \cite{FM} the number of colors needed was at least $n^\a$ for $\a=\a(k)$, in this paper we reduce the number of colors to logarithmic in $n$.

Large parts of the proofs in \cite{FM} are still relevant and so we will quote them in place of re-proving them. We have realised that some minor simplifications are possible. So if the proofs in \cite{FM} need to be tweaked, we will indicate what is needed in an appendix.

Before formally stating our theorem we will define the Glauber dynamics.
All of the aforementioned results on coloring graphs (except Vigoda \cite{vigoda})
analyze the Glauber dynamics, which is a simple and popular Markov
chain for generating a random $q$-coloring.

Let $\cQ$ denote the set of proper $q$-colorings of $H$. For a coloring $X\in\cQ$ we define
$$B(v,X)=\set{c\in [q]:\;\exists e\ni v\ \text{such that}\ X(x)=c\ for\
all\ x\in e\setminus\set{v}}$$
be the set of colors unavailable to $v$.

Then let $Q=\{1,2,\dots,q\}$ and
$$A(v,X)=Q\setminus B(v,X).$$
For technical purposes, the state space of the Glauber dynamics is $\Omega=Q^V\supseteq\cQ$.
From a coloring $X_t\in\Omega$, the evolution $X_t\rightarrow X_{t+1}$
is defined as follows:

{\bf Glauber Dynamics}\vspace{-.2in}
\begin{description}
\item[(a)] Choose $v=v(t)$ uniformly at random from $V$.
\item[(b)] Choose color $c=c(t)$ uniformly at random from $A(v,X_t)$.
If $A(v,X_t)$ is empty we let $X_{t+1} = X_t$.
\item[(c)] Define $X_{t+1}$ by
$$X_{t+1}(u)=\begin{cases}X_t(u)&u\neq v\\c&u=v\end{cases}$$
\end{description}
We will assume from now on that
\beq{qd}{
q\leq 2\D
}
If $q>2\D$ then we defer to Jerrum's result \cite{Jerrum}.

Let $Y$ denote a coloring chosen uniformly at random from $\cQ$. We will prove the following:
\begin{theorem}\label{th1}
Let $H$ be a $k$-uniform simple hypergraph with maximum degree $\D$
where $k\geq 3$.
Suppose that \eqref{qd} holds and that
\beq{vq}{
q\geq \max\set{C_k\log n,10k\e^{-1}\D^{1/(k-1)}},
}
where $C_k$ is sufficiently large and depends only on $k$ and 
\beq{defeps}{
\e=\frac{1}{50k^2}.
}
Suppose that the initial coloring $X_0$ is chosen randomly from $\Omega$.
Then for an arbitrary constant $\d>0$ we have
\beq{rapid}{
d_{TV}(X_t,Y)\leq \d
}
for $t\geq t_\d$, where
$t_\d=2n\log(2n/\d)$.
\end{theorem}
\begin{remark}
The definition of $\e$ has been changed from \cite{FM}. It results in slightly better bounds for $q$. It would seem that we could absorb it into the expression \eqref{vq}, but its value is used throughout the paper.
\end{remark}
Here $d_{TV}$ denotes variational distance i.e.
$\max_{S\subseteq \cQ}|\Pr(X_t\in S)-\Pr(Y\in S)|$.

Note that we do not claim rapid mixing from an arbitrary start.
Indeed, since we are using relatively few colors, it is possible
to choose an initial coloring from which there is no Glauber move i.e.
we do not claim that the chain is ergodic, see \cite{FM} for examples of blocked colorings.

The algorithm can be used in a standard way, \cite{Jerrum},
to compute an approximation to the number of proper colorings of $H$.

We can also prove the following. We can consider Glauber Dynamics
as inducing a graph $\G_{\cQ}$ on $\cQ$ where two colorings are connected
by an edge if there is a move taking one to the other. Note that if Glauber
can take $X$ to $Y$ in one step, then it can take
$Y$ to $X$ in one step.
\begin{corollary}\label{cor1}
The graph $\G_{\cQ}$ contains a giant component $\cQ_0$ of size $(1-o(1))|\cQ|$.
\end{corollary}

\section{Good and bad colorings}
Let $X\in \Omega$ be a coloring of $V$. For a vertex $v\in V$ and $1\leq i\leq k-1$ let
$$E_{v,i,X}=\set{e:\;v\in e\ and\ |\set{X(w):w\in e\setminus\set{v}}|=i}$$
be the set of edges $e$ containing $v$ in which $e\setminus\set{v}$
uses exactly $i$ distinct colors under $X$.
Let $y_{v,i,X}=|E_{v,i,X}|$ so that
$|B(v,X)|\leq y_{v,1,X}$ for all $v,X$. Let $\e$ be as in \eqref{defeps}. We define the sequence $\bse=\e,\e^2,\ldots,\e^{k-2}$.
\begin{definition}\label{def2}
We say that $X$ is {\em $\bse$-bad} if $\exists v\in V, 1\leq i\leq k-2$
such that
\beq{m1}{
y_{v,i,X}\geq \m_i\text{ where }\m_i=
(\e q)^i.
}
Otherwise we say that $X$ is {\em $\bse$-good}. 
\end{definition}
\begin{remark}
In \cite{FM}, $\m_i$ is the minimum of $(\e q)^i$ and a more complicated term. This second term is no longer needed.
\end{remark}
Given Definition \ref{def2}, we have
\beq{mudef}{
10k\m_i\leq \m_{i+1}\le\e q\m_i\qquad for\ 1\leq i\leq k-3.
}
It is convenient to define
\beq{mudef0}{
\m_{k-1}=\D.
}

Note that if $X$ is $\bse$-good then $|A(v,X)|\geq (1-\e)q$ for all $v\in V$.

In this section we will show that almost all colorings of $\Omega$ are
$\bse$-good and almost all colorings in $\cQ$ are $\bse$-good. This is where we are able to improve our results over \cite{FM}. 

Consider a random coloring $X\in\Omega$. For a vertex $v\in V$ we let $\cA_v=\cA_\e(v)$ denote the event $\set{v\text{ is not $\bse$ good}}$. For an edge $e\in E$ we let $\cB_e$ denote the event $\set{e\ \text{is not properly colored})}$. 

Let $\Pr_\Omega$ indicate that the random choice is from $\Omega$ and let $\Pr_\cQ$ indicate that the random choice is from $\cQ$. 

Now consider a dependency graph, in the context of the local lemma. The
events are $\cB_e,e\in E$ and $\cB_e$ and $\cB_f$ are independent if $e\cap f=\emptyset$. Note that for $v\in V$ the event $\cA_v$ is  independent of events not in $\cN_v$, where $$\cN_v=\set{f:e\cap f\neq \emptyset\text{ for some }e\ni v}.$$

Fix an edge $e\in H$. Clearly, 
$$p=\Pr_\Omega(\cB_e)=\frac{1}{q^{k-1}}.$$
We will next choose $x_e,e\in E$ to satisfy
\beq{vp}{
p\leq x_e\prod_{f\in E,f\cap e\neq \emptyset}(1-x_f).
}
We choose
\beq{xv}{
x_e=\th_q=\frac{2}{q^{k-1}}\leq \frac12\text{ for }e\in E.
}
Then, using \eqref{vq} and $(1-x)\geq e^{-x/(1-x)}$ for $0<x<1$ we see that
\beq{}{
x_e\prod_{f\in E,f\cap e\neq \emptyset}(1-x_f)\geq \th_q\exp\set{-\frac{k\D\th_q}{1-\th_q}}
\geq \th_qe^{-2k\D\th_q}
=\frac{2}{q^{k-1}}\exp\set{-\frac{4k\D}{q^{k-1}}}
\geq \frac{2e^{-1/2}}{q^{k-1}}
\geq p.
}
This verifies \eqref{vp}. Theorem 2.1 of Haeupler, Saha and Srinivasan \cite{HSS} then implies that for $v\in V$ we have
\beq{LLL}{
\Pr_\cQ(\cA_v)\leq \Pr_\Omega(\cA_v)\prod_{f\in \cN_v}(1-x_f)^{-1}.
}
This theorem is the basis of our improvement. As stated in \cite{HSS}, there is a short easy proof of this and for completeness we give an outline in an appendix.

Now \cite{FM} proves that 
\beq{upper}{
\Pr_\Omega(\cA_v)\leq e^{-\e q}.
}
\begin{remark}
The defintion of $\e$ has changed from \cite{FM} and so we feel obliged to verify \eqref{upper} in an appendix.
\end{remark}
So, \eqref{LLL} implies that
\beq{}{
\Pr_\cQ(\exists v\in V:\cA_v)\leq ne^{-\e q}\prod_{f\in \cN_v}(1-x_f)^{-1}\leq n\exp{\set{-\e q+\frac{4k\D}{q^{k-1}}}}\leq ne^{\frac12-\e q}=o(1),
}
since $q\geq 2\e^{-1}\log n$.

Thus w.h.p., a $q$-coloring chosen uniformly at random from either $\Omega$ or from $\cQ$ is $\bse$-good.
\section{Persistence of goodness}
The following two lemmas are proved in \cite{FM}:
\begin{lemma}
\beq{first}{
\Pr(X_t\ is\ 2\bse-good\ for\ t\leq t_0\mid X_0\ is\ \bse-good)\geq 1-2^{-\m_1/2}.
}
where
\beq{t0}{
t_0=\frac{n}{4k^2e}.
}
\end{lemma}
\begin{lemma}\label{cont}
\beq{step2}{
\Pr(X_{t_0}\ is\ \bse-good\mid X_0\text{ is }\bse-good)\geq 1-e^{-c\m_1}\qquad for\ some\ c>0.
}
\end{lemma}
The constant $c$ in \eqref{step2} depends only on $k$. Part of the proof of Lemma \ref{cont}, involves the inequality $\frac{4\e k^2}{1-2\e}\leq \frac{1}{10}$, see (26) of \cite{FM}. Our choice of $\e$ satisfies the latter inequality.
\section{Coupling Argument}
Now consider a pair $X,Y$ of copies of our Glauber chain. Let
$$h(X_t,Y_t)=|\set{v\in V:X_t(v)\neq Y_t(v)}|$$
be the Hamming distance between $X_t,Y_t$. The paper \cite{FM} describes a simple coupling
between the chains and shows that
that
\beq{h1}{
\E(h(X_{t+1},Y_{t+1})\mid X_t,Y_t)\leq \brac{1-\frac{1}{2n}}h(X_t,Y_t)
} 
if $X_t,Y_t$ are both $2\bse$-good.

Summarising, we have that with probability at least $1-e^{-c\m_1}$
for some positive constant $c$, we have that both of $X_0$ and $Y_0$ are $\bse$-good,
both $X$ and $Y$ are $2\bse$-good for $t_0$ steps and both of $X_{t_0}$ and $Y_{t_0}$ are $\bse$-good. If we run the chain for $t_0t^*$ steps, where
$t^*=e^{c\m_1/2}$ then the probability
that either chain stops being $2\bse$-good is at most $t^*e^{-c\m_1}= e^{-c\m_1/2}$.
Conditional on these events, $\E(h(X_{t_\d},Y_{t_\d})\leq \d/2$ and together with the fact that
the variation distance between $X_t$ and $Y_t$ is monotone non-increasing, this
implies \eqref{rapid}.
(Note that $\d$ includes the probability that either of $X_0,Y_0$ are not $2\bse$-good).
Choosing $C_k$ large enough so that $c\e C_k=\frac{cC_k}{50k^2}\geq 4$ completes the proof of Theorem \ref{th1}. (This choice of $C_k$ ensures that $t^*\geq n^2$.)
\subsection{Proof of Corollary \ref{cor1}}
The proof of Theorem \ref{th1} shows that if $X,Y\in\cQ$ are both $\bse$-good
then there is a path from $X$ to $Y$ in $\cQ$
of length $O(n\log n)$. Since almost all of $\cQ$ is $\bse$-good, we are done.
\proofend

\appendix
\section{Proof of \eqref{LLL}}
\begin{align}
\Pr_\cQ(\cA_v)&=\Pr_{\Omega}(\cA_v\mid \bigcap_{e\in E}\cB_e)\\
&=\frac{\Pr_{\Omega}(\cA_v\cap \bigcap_{e\in \cN_v}\cB_e\mid \bigcap_{e\notin \cN_v}\cB_e)} {\Pr_{\Omega}(\bigcap_{e\in \cN_v}\cB_e\mid \bigcap_{e\notin \cN_v}\cB_e)}\\
&\leq \frac{\Pr_{\Omega}(\cA_v\mid \bigcap_{e\notin \cN_v}\cB_e)} {\Pr_{\Omega}(\bigcap_{e\in \cN_v}\cB_e\mid \bigcap_{e\notin \cN_v}\cB_e)}\\
&=\frac{\Pr_{\Omega}(\cA_v)} {\Pr_{\Omega}(\bigcap_{e\in \cN_v}\cB_e\mid \bigcap_{e\notin \cN_v}\cB_e)}\\
&\leq \frac{\Pr_\Omega(\cA_v)}{\prod_{e\in \cN_v}(1-x_e)}. \label{last}
\end{align}
Inequality \eqref{last} follows from a standard proof of the Lov\'asz Local Lemma, see for example Alon and Spencer \cite{AS}, 3rd Edition, (5.4).
\section{Proof of \eqref{upper}}
The proof in \cite{FM} starts with \eqref{upper1}:
\begin{align}
\Pr(y_{v,i,X}\geq \m_i)
&\leq \bfrac{e\binom{k-1}{i}(i/q)^{k-1-i}\D}{\m_i}^{\m_i}\label{upper1}\\
&=\bfrac{e\binom{k-1}{i}(i/q)^{k-1-i}\D}{(\e q)^i}^{\m_i}\\
&\leq \bfrac{k^ie^{i+1}i^{k-1}\D}{i^{2i}(\e q)^{k-1}}^{\m_i}\\
&\leq \bfrac{k^ie^{i+1}i^{k-1}}{i^{2i}(500k^2)^{k-1}}^{\m_i}\\
&\leq 10^{-2\m_i}\\
&\leq 10^{-2\e q}.\label{lastone}
\end{align}
Considering the union of the $k-2$ events that define $\cA_v$, we inflate the bound in \eqref{lastone} by $k-2$ and obtain \eqref{upper}. 

\begin{thebibliography}{99}
\bibitem{AS} N. Alon and J. Spencer, The Probabilistic Method, 3rd Edition, John Wiley and Sons, Hoboken NJ, 2008.

\bibitem{BW}
G. R. Brightwell\ and\ P. Winkler,
Random colorings of a Cayley tree.
{\em Contemporary combinatorics}, 10:247--276, 2002.

\bibitem{CV} B. Cousins and S. Vempala, Bypassing KLS: Gaussian Cooling and an $O^*(n^3)$ Volume Algorithm, STOC 2015.

\bibitem{DF}
M. Dyer, A. Frieze.
\newblock Randomly coloring graphs with lower bounds
on girth and maximum degree.
{\em Random Structures and Algorithms}, 23(2):167-179, 2003.

\bibitem{DFK} M.E. Dyer, A.M. Frieze and R. Kannan,
A random polynomial time algorithm for approximating the volume of convex
bodies, {\em Journal of the Association for Computing Machinery}
38(1):1--17, 1991.

\bibitem{FM} A.M. Frieze and P. Melsted, Randomly coloring simple hypergraphs, {\em Information Processing Letters} 11 (2011) 848-853.

\bibitem{FM1} A.M. Frieze and D. Mubayi, Colouring Simple Hypergraphs, {\em Journal of Combinatorial Theoty B} 103 (2013) 767-794.

\bibitem{FriVig} A.M. Frieze and E. Vigoda, A survey on the use of Markov
chains to randomly sample colorings,
in {\em Combinatorics, Complexity and Chance, A tribute to Dominic Welsh},
(G. Grimmett, C. McDiarmid Eds.) (2007) 53-71.

\bibitem{HSS} B. Haeupler, Saha and Srinivasan, New Constructive Aspects of the Lov\'asz Local lemma, {\em Journal of the ACM} 58 (2011) .

\bibitem{Jerrum} M.R. Jerrum,  A very simple algorithm for estimating
the number of $k$-colorings of a low-degree graph,
{\em Random Structures and Algorithms}, 7(2):157--165, 1995.

\bibitem{JSV} M.R. Jerrum, A. Sinclair and E. Vigoda,
A polynomial-time approximation algorithm for the permanent of a
matrix with non-negative entries,
{\em Journal of the Association for Computing Machinery}, 51(4):671-697, 2004.

\bibitem{KLS} R. Kannan, L. Lov\'asz and M. Simonovits,
Random walks and an $O^*(n^5)$ volume algorithm for convex bodies,
{\em Random Structures and Algorithms}, 11(1):1--50, 1997.

\bibitem{LV} L. Lov\'asz and S. Vempala,
Simulated Annealing in Convex Bodies and an $O^*(n^4)$
Volume Algorithm.
In \emph{Proceedings of the 44th Annual IEEE Symposium on Foundations of
Computer Science} (FOCS), 650-659, 2003.

\bibitem{SS} J. Salas and A. Sokal,
Absence of phase transition for antiferromagnetic Potts models via the Dobrushin
uniqueness theorem,
{\em Journal of Statistical Physics}, 86(3-4):551--579, 1997.

\bibitem{vigoda} E. Vigoda,
Improved bounds for sampling colorings,
{\em Journal of Mathematical Physics}, 41(3):1555-1569, 2000.



\end{thebibliography}
\end{document}